\documentclass[aps,twocolumn,showpacs]{revtex4}
\usepackage{dcolumn}
\usepackage{float}
\usepackage{graphicx}
\usepackage{amsmath}
\usepackage{verbatim}
\usepackage{amssymb}
\usepackage{psfrag}
\usepackage{color}
\usepackage{bm}
\usepackage{wrapfig}
\usepackage{subfigure}
\usepackage{makeidx}
\usepackage{bm}
\usepackage{epsf}
\usepackage{multirow}
\usepackage{mathrsfs}
\usepackage{braket}
\usepackage[colorlinks,linkcolor=blue,urlcolor=blue,citecolor=blue]{hyperref}

\begin{document}
\title{A corresponding relationship between nonlinear Hermitian systems and linear non-Hermitian models}
\author{Yu-Hao Wang$^{1}$}
\author{Yan-Hong Qin$^{2,3}$}
\author{Jie Liu$^{4}$ }
\author{Li-Chen Zhao$^{1,5,6}$}\email{zhaolichen3@nwu.edu.cn}

\address{$^{1}$School of Physics, Northwest University, Xi'an, 710127, China}
\address{$^{2}$College of Mathematics and System Sciences, Xinjiang University, Urumqi, China, 830046}
\address{$^{3}$Institute of Mathematics and Physics, Xinjiang University, Urumqi, China, 830046}
\address{$^{4}$Graduate School, China Academy of Engineering Physics, Beijing 100193, China}
\address{$^{5}$NSFC-SPTP Peng Huanwu Center for Fundamental Theory, Xi'an 710127, China}
\address{$^{6}$Shaanxi Key Laboratory for Theoretical Physics Frontiers, Xi'an 710127, China}

\begin{abstract}
We note that the non-orthogonality of states and their coincidence at the degeneracy point are both admitted by nonlinear Hermitian systems and linear non-Hermitian systems. These striking characteristics motivate us to re-investigate the localized waves of nonlinear Hermitian systems and the eigenvalue degeneracies of linear non-Hermitian models, based on several well-known Lax-integrable systems that have wide applications in nonlinear optics. We choose nonlinear Schr\"odinger equation integrability hierarchy to demonstrate the quantitative relations between dynamics of nonlinear Hermitian systems and eigenvalue degeneracies of linear non-Hermitian models. Specifically, the degeneracies of the real or imaginary spectrum of the linear non-Hermitian matrices are uncovered to clarify several essential characteristics of nonlinear localized waves, such as breathers, rogue waves, and solitons. We find that the exceptional points generally correspond to rogue waves for modulational instability cases and dark solitons with maximum velocity for the modulational stability cases. These insights provide another interesting perspective for understanding nonlinear localized waves, and hint that there are closer relations between nonlinear Hermitian systems and linear non-Hermitian systems.
\end{abstract}

\date{\today}
\maketitle

\section{Introduction}
Linear non-Hermitian systems have received intense attention recently, due to their striking characteristics, such as the non-orthogonality of states and their coincidence at the degeneracy points \cite{book1,book2,book3}. These properties indeed induce many novel physical phenomena \cite{rmp1,ap1,prl2,np1,np2}, such as nontrivial topology \cite{sc1},  enhanced sensitivity near spectral degeneracies\cite{nature1},  and non-Hermitian skin effects \cite{prl1}. Interestingly, we note that localized wave states of nonlinear Hermitian systems can also admit the non-orthogonality characteristics and they also coincide at the degeneracy point of soliton excitation energies. These characteristics can be seen clearly in the two mass branches of soliton solutions in nonlinear coupled systems \cite{zhao20,meng22,gaox25}. The two aspects hint that there should be some intrinsic relations between nonlinear Hermitian systems and linear non-Hermitian systems.

It should be noted that the Lax-pairs are generally linear non-Hermitian matrices for nonlinear Hermitian systems admitting Lax-integrable properties \cite{darboux1,darboux2,JYang}, which provides possibilities to discuss the relationship between them preliminarily. Although the Lax-integrable systems are well known in mathematical physics, the potential relationships between the solutions and spectrum characteristics of the non-Hermitian models were usually overlooked, and most mathematicians and physicists mainly solved the Lax-pairs and constructed various localized wave solutions of nonlinear systems \cite{ds1,ds2,zhaolax,dd6}. This oversight was partly due to a limited understanding of degeneracies in non-Hermitian systems. Recent developments of non-Hermitian physics \cite{cp1,np4,nrp1} provide some new perspectives and inspire us to discuss the potential relationships.

In this work, we choose the nonlinear Schr\"odinger (NLS) equation integrability hierarchy to uncover quantitative relations between the degeneracies of linear non-Hermitian matrices and the localized wave solutions of nonlinear Hermitian systems, considering the hierarchy has wide applications in physical systems \cite{BECbook,nonliearopbook,financialbook} and many experimental realizations \cite{experiment1,experiment2,experiment3,experiment4}.
We analyze the eigenvalue degeneracy of Lax-pairs in details. The eigenvalues can be used to construct families of nonlinear localized waves. For simplicity and without loss of generality, we mainly discuss the localized waves on plane wave backgrounds. The degeneracies of the real or imaginary spectrum of the linear non-Hermitian matrices are uncovered to clarify several essential characteristics of the usual nonlinear localized waves, such as breathers \cite{kmexp,abexp1,abexp2}, rogue waves \cite{rwexp}, dark solitons \cite{dsexp}, and bright solitons \cite{bsexp}, which have been investigated widely in nonlinear optical systems \cite{nos1,nos2,nos3}. In particular, we find that exceptional points coincide with rogue waves for modulational instability cases, and dark solitons possessing the maximum speed for the modulational stability cases.
These insights offer a novel perspective for understanding the relations between nonlinear Hermitian and linear non-Hermitian systems.

The structure of the paper is as follows. In section \ref{the scalar Manakov system}, we consider the NLS equation integrability hierarchy to analyze the mapping between the spectral degeneracies (real or imaginary) of the linear non-Hermitian matrices and nonlinear localized waves. In section \ref{the Coupled Manakov system}, we extend the coupled NLS equation integrability hierarchy, reveal the mapping between the degeneracies of linear non-Hermitian matrices and the vector nonlinear localized waves, and further visualize the distribution of degeneracies in the parameter space.
Finally, in section \ref{conclusion}, we summarize and discuss the results.

\section{the scalar NLS system}\label{the scalar Manakov system}

\begin{figure}[t]
\begin{center}
\subfigure{\includegraphics[width=90mm]{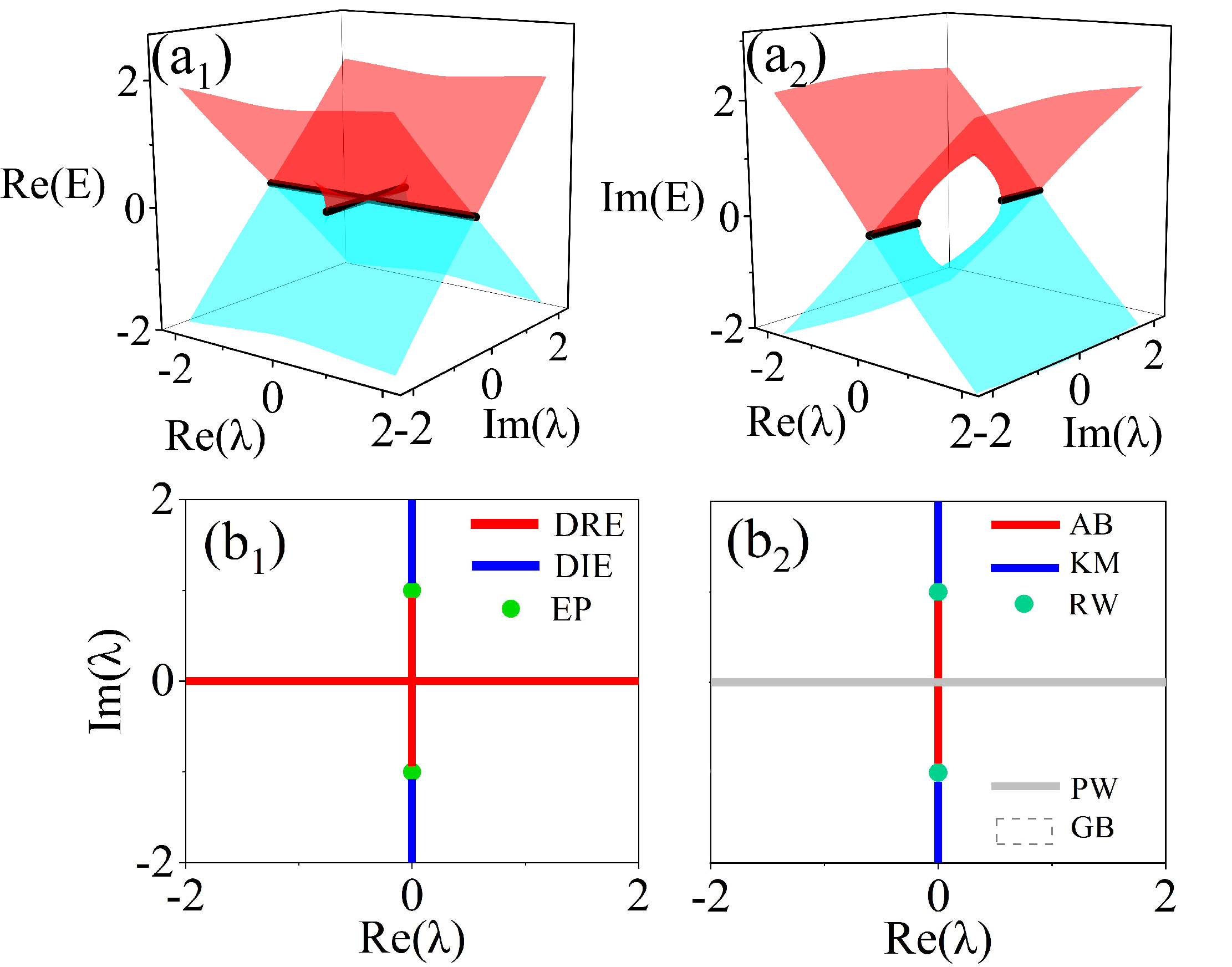}}
\end{center}
\caption{(a$_1$) and (a$_2$) The real and imaginary parts of the energy eigenvalues near EPs for $\sigma=1$.
The red and blue surfaces represent the positive and negative energy branches, respectively, black lines represent energy degeneracy.
(b$_1$) The corresponding degeneracy regions near the EPs, where the red lines, blue lines and green dots represent DRE, DIE and EPs, respectively.
(b$_2$) The soliton existence regions on a plane wave background.
The red, blue, and gray lines correspond to AB, KM, and PW solutions, respectively, while green dots denote RW solutions.
The blank regions represent GB solutions.
Parameters are $s=1$ and $\sigma=1$.}\label{NLSg1}
\end{figure}
\begin{figure}[t]
\begin{center}
\subfigure{\includegraphics[width=87mm]{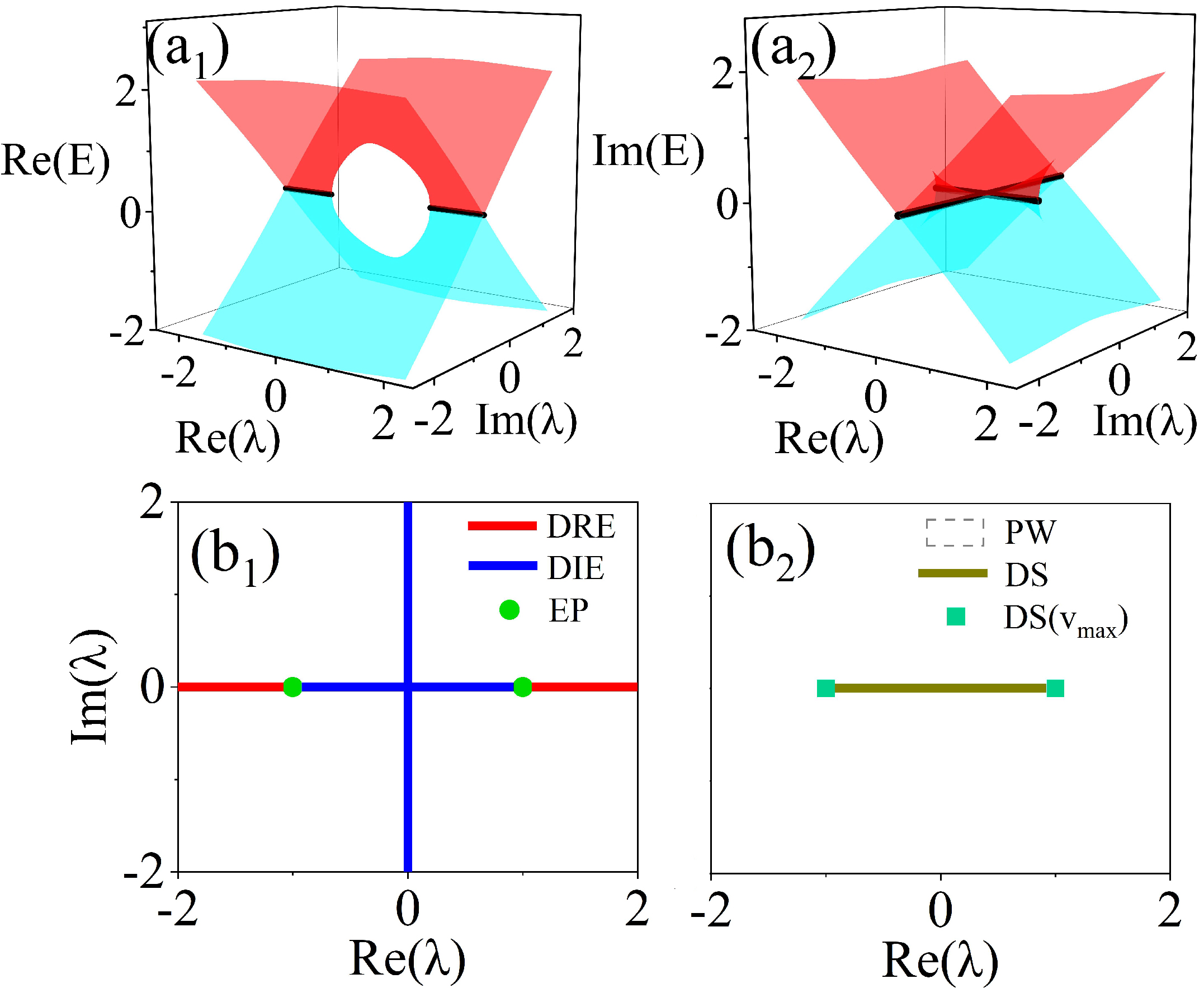}}
\end{center}
\caption{(a$_1$) and (a$_2$) The real and imaginary parts of the energy eigenvalues near EPs for $\sigma=-1$.
The red and blue surfaces represent the positive and negative energy branches, respectively, black lines represent energy degeneracy.
(b$_1$) The corresponding degeneracy regions near the EPs, where the red lines, blue lines and green dots represent DRE, DIE and EPs, respectively.
(b$_2$) The soliton existence regions on a plane wave background, where olive green line and green squares represent DS and DS solutions at maximum velocity, respectively.
The blank regions represent PW solutions.
Other parameters: $s=1$.
}\label{NLSg-1}
\end{figure}

The NLS equation hierarchy, a class of well-known nonlinear Hermitian systems with Lax-integrable properties, has been widely used to describe nonlinear wave dynamics in various physical systems
\cite{BECbook,nonliearopbook,financialbook,experiment1,experiment2,experiment3,experiment4}, including nonlinear optical fiber \cite{experiment1}, Bose-Einstein condensates \cite{bec}, plasma systems \cite{experiment3}, and water wave tanks \cite{experiment2}.  We begin with one of the simplest models--the scalar NLS equation--to explore the quantitative relationships between the degeneracies of linear non-Hermitian matrices and the localized wave solutions of nonlinear Hermitian systems.
In nonlinear optical systems, the propagation of nonlinear waves is governed by an equation that is related to the NLS:

\begin{eqnarray}\label{NLS}
i\frac{\partial \psi(\tau,\xi)}{\partial \xi}+\frac{1}{2}\frac{\partial^2\psi(\tau,\xi)}{\partial \tau^2}+\sigma|\psi(\tau,\xi)|^2\psi(\tau,\xi)=0,
\end{eqnarray}
where $\psi(\tau,\xi)$ represents the wavefunction, with $\tau$ and $\xi$ corresponding to the retarded time and the normalized distance along the fiber, respectively. The coefficient $\sigma$ is the nonlinearity parameter that determines the strength of interactions, governing the Kerr effect in nonlinear optical systems.
The above equation is Lax-integrable, and various nonlinear localized wave solutions have been derived through methods such as the Darboux transformation \cite{darboux1,darboux2,darboux5}, the Hirota bilinear method \cite{Hirota2}, and the inverse scattering transform \cite{inverse}.  It admits the following Lax pair \cite{NLSlax}:
\begin{subequations}\label{spectral problems}
\begin{align}
&\Phi_\tau=U\Phi,\\
&\Phi_\xi=V\Phi,
\end{align}
\end{subequations}
with
\begin{subequations}\label{UV}
\begin{align}
 U&=i\begin{pmatrix}
    \lambda &  \sigma \psi^* \\
    \psi & -\lambda\\
\end{pmatrix},\\
V&=\begin{pmatrix}
    i\lambda^2-\frac{1}{2}i\sigma|\psi|^2 & \sigma(\frac{1}{2}\psi_\tau^*+i\lambda\psi^*) \\
    -\frac{1}{2}\psi_\tau+i\lambda\psi & -i\lambda^2+\frac{1}{2}i\sigma|\psi|^2\\
\end{pmatrix},
\end{align}
\end{subequations}
where the symbol ``*" denotes the complex conjugate, and $\lambda$ is a complex spectral parameter. Clearly, $U \neq U^\dagger$ and $V \neq V^\dagger$ (here $\dagger$ denotes the Hermitian conjugate), confirming that both matrices are non-Hermitian. It can thus be properly considered as Hamiltonians for a non-Hermitian system, allowing us to study degeneracies in its Lax-pairs' eigenvalue spectrum. As is well established, nonlinear localized wave solutions in the corresponding Hermitian system can be obtained through Lax-pair. This provides a framework for investigating the correspondence between nonlinear localized waves and the eigenvalue degeneracy of non-Hermitian matrices.

Since we consider localized wave solutions of breathers, rogue waves and solitons  on the non-vanishing background, we consider a plane-wave seed solution of the form $\psi = s e^{{i} s^2 \sigma \xi}$ to solve the above Lax-pair, where $s$ is the amplitude of the plane wave background.
By introducing a transformation matrix $P$ defined as $P = \mathrm{diag}(e^{i s^2 \sigma \xi / 2},e^{-i s^2 \sigma \xi / 2})$, the variable-coefficient differential equation can be converted into a constant-coefficient equation, yielding  $(P\Phi)_{\tau} = U_0(P\Phi)$ and $(P\Phi)_{\xi} = V_0(P\Phi)$. The constant matrices $U_0$ and $V_0$ are given by:
\begin{subequations}\label{U0V0}
\begin{align}
U_0&=i\begin{pmatrix}
    \lambda & s \sigma \\
    s & -\lambda\\
\end{pmatrix},\\
V_0&=\lambda U_0.
\end{align}
\end{subequations}
Since a linear relationship exists between the non-Hermitian matrices $U_0$ and $V_0$, $U_0$ can serve as the effective Hamiltonian of a non-Hermitian system, whose eigenstates satisfy the eigenvalue equation:
\begin{eqnarray}\label{Ephi}
i\begin{pmatrix}
    \lambda & s \sigma \\
    s & -\lambda\\
\end{pmatrix}\varphi=E\varphi.
\end{eqnarray}
The eigenfunction $\varphi$ corresponds to a given set of these parameters. The energy eigenvalues are given by:
\begin{eqnarray}\label{EE}
E_{\pm}=\pm\sqrt{-s^2\sigma-\lambda^2}.
\end{eqnarray}
Taking the square root of a complex function induces a one-to-two correspondence between the spectral parameter $\lambda$ and the eigenvalues $E_{\pm}$. The two energy branches generate three types of degeneracies: double degeneracy in the real part of eigenvalues (DRE), double degeneracy in the imaginary part of eigenvalues (DIE), and an energy degeneracy point where both eigenvalues become equal to zero. Naturally, at energy degeneracy point, the Hamiltonian $U_0$ becomes non-diagonalizable, and the eigenvectors of the two energy branches become non-orthogonal and coincide. This phenomenon, where both eigenvalues and eigenvectors coalesce in a non-Hermitian system, is known as an exceptional point  (EP) \cite{ep1,ep2,nrp1}. Here, these EPs emerge from two coalescing eigenstates and are thus termed square-root EPs or second-order EPs (2-EPs). Then, the degeneracy landscape can be systematically explored through variations of the nonlinearity parameter $\sigma$, background amplitude $s$ and the spectral parameter $\lambda$.

\begin{figure}[t]
\begin{center}
\subfigure{\includegraphics[width=88mm]{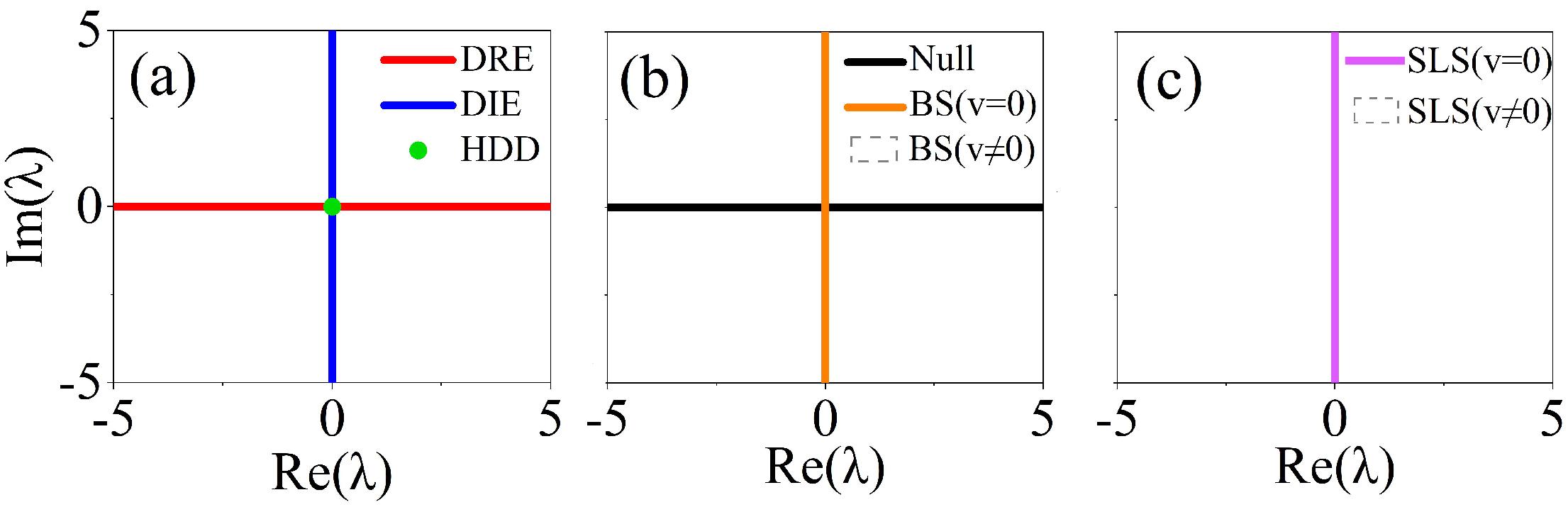}}
\end{center}
\caption{(a) Degeneracy regions for zero background, which are independent of the nonlinear parameter $\sigma$.
The red, blue lines and green dot represent DRE, DIE and HDD, respectively.
(b) Soliton existence regions for $\sigma = 1$. The black and orange line represent zero solution and stationary BS, respectively.
The blank regions represent BS with velocity.
(c) Soliton existence regions for $\sigma = -1$. The purple line denotes stationary SLS.
The blank regions represent SLS with velocity.
}
\label{NLSs0}
\end{figure}

We show the real and imaginary parts of the energy eigenvalues near EPs in Fig.~\ref{NLSg1}(a$_1$)-(a$_2$) for $\sigma=1$ and Fig.~\ref{NLSg-1}(a$_1$)-(a$_2$) for $\sigma=-1$,  with a nonzero background  $s=1$. It is clear that the geometry of the eigenvalue surface is rather unusual, as it does not resemble any two-dimensional surface embeddable in three-dimensional Euclidean space. Remarkably, the two eigenvalues remain a smooth connection even as the system crosses a branch cut continuously. The corresponding degeneracy regions near the EPs are visualized  via Re($\lambda$) and  Im($\lambda$) in Figs.~\ref{NLSg1}(b$_1$) and~\ref{NLSg-1}(b$_1$), respectively, where green dots, red lines and blue lines represent 2-EPs, DRE and DIE regions, respectively. A striking difference is observed between the degeneracy regions in the non-Hermitian model Eq.~\eqref{Ephi} for $\sigma=1$ and $\sigma=-1$.

For the case of zero background $s=0$,  the energy eigenvalues simplify to $E_{\pm}=\pm i \sqrt{\lambda^2}$,  which are independent of the nonlinear parameter $g$. The regions of eigenvalue degeneracy in the Re($\lambda$)-Im($\lambda$)  plane are shown in  Fig.~\ref{NLSs0}(a). It is worth noting that the $U_0$ becomes a zero matrix when $\lambda = 0$,  converting the originally non-Hermitian Hamiltonian into a Hermitian one. In this case, a Hermitian double degeneracy (HDD) emerges, represented by the green dot in  Fig.~\ref{NLSs0}(a),  distinct from the EP in the non-Hermitian case. The corresponding eigenvectors at the HDD point maintain both completeness and orthogonality.

The degeneracy of eigenvalues in linear non-Hermitian models arising from Lax pair is well understood. It is well known that the Lax pair can be widely  employed to construct various types of nonlinear localized wave solutions and plane wave solutions (PWs), including  Kuznetsov-Ma breathers (KMs) \cite{km1,km2}, Akhmediev breathers (ABs) \cite{ab1,ab2,ab3}, general breathers (GBs) \cite{gb}, rogue waves (RWs) \cite{rw}, bright solitons (BSs), dark solitons (DSs) \cite{ds1,ds2,inverse} and singular localized states (SLSs). Moreover, the eigenvalues of the Lax pair play a crucial role in generating specific types of nonlinear localized waves \cite{NLSlax,dbkm,dbab,dbrw1,dd6}. The existence regions of various localized wave solutions are determined by eigenvalue-dependent criteria. Then, these solution regimes can be directly mapped in the Re($\lambda$)-Im($\lambda$)  plane. Therefore, we can analyze  the corresponding  relations between localized wave solutions of nonlinear Hermitian systems and eigenvalue degeneracies of linear non-Hermitian models.

For $\sigma=1$, the Hermitian system \eqref{NLS} possess  modulation instability and supports KM, AB, GB and RW solutions on a plane wave background.  A RW emerges when the two eigenvalues are equal \cite{NLSlax,dbrw1}, AB form when the eigenvalues are complex conjugates of each other \cite{dbab}, KM breathers form when the imaginary parts of eigenvalues are equal and the real parts are opposite of each other \cite{dbkm}. Based on the existence conditions determined by their eigenvalues, we present the existence regions of these solutions in the Re($\lambda$)-Im($\lambda$)  plane in Fig.~\ref{NLSg1}(b$_2$). It can be observed that the boundary between KM and AB solutions lies at $\lambda = \pm i$, coinciding with the RW solution---a known limiting case of KM/AB solutions as the period approaches infinity. A remarkable correspondence emerges from the comparative analysis of Fig.~\ref{NLSg1}(b$_1$) and Fig.~\ref{NLSg1}(b$_2$): (i) the AB solution domain in the nonlinear Hermitian system precisely coincides with the dispersive radiative excitation DRE regime in the non-Hermitian model (red lines); (ii) the KM solution regime directly maps onto the DIE domain (blue lines); (iii) the GB solutions have no degeneracies;  (iv) while RW solution emerges at the EPs position in the non-Hermitian framework (green dots). When the background amplitude $s=0$,  the parameter space only exists BS solutions, the stationary BS indicated by the orange lines in Fig.~\ref{NLSs0}(b), corresponding to the DIE regime represented by blue lines in Fig.~\ref{NLSs0}(a). Additionally, a BS at zero density ($s=0$, $\lambda=0$)  coincides with the HDD point,  as marked by the green dot in Fig.~\ref{NLSs0}(a).

For $\sigma=-1$, there is no modulation instability in the Hermitian system \eqref{NLS}, which admits the DS and PW solutions, as shown in Fig.~\ref{NLSg-1}(b$_2$). It should be pointed out that the DS solutions acquire a series of real spectral parameters \cite{dd6}. The boundary between the DS and PW solutions occurs at $\lambda = \pm 1$, where the DS attains its maximum velocity. Notably, this maximum velocity matches the condensate $c_s$  in Bose-Einstein condensate systems ($c_s=1$).  Compared with the eigenvalue degeneracy of the non-Hermitian model  shown in Fig.~\ref{NLSg-1}(b$_1$), we find that: (i) when the DS velocity is below its maximum, the DS solution in the nonlinear Hermitian system corresponds to the DIE in the non-Hermitian model; (ii) at maximum velocity, the DS solution aligns with the EPs in the non-Hermitian model; and (iii) when the DS velocity exceeds the sound speed, it degenerates into a plane wave solution, corresponding to the DRE regime.
For the $s=0$ case, only SLS solutions appear, the stationary SLS as shown in Fig.~\ref{NLSs0}(c), which directly map to the DIE regime shown in Fig.~\ref{NLSs0}(a).

These results demonstrate that eigenvalue degeneracies in the real or imaginary spectrum of linear non-Hermitian matrices can effectively and quantitatively characterize nonlinear localized wave solutions in scalar Hermitian systems. The coupled NLS equation hierarchy admits even more abundant vector localized wave solutions, whose relationship to eigenvalue degeneracies in non-Hermitian matrices is expected to be richer than in the scalar case and warrants further  investigation.

\section{the Coupled NLS system}\label{the Coupled Manakov system}

The coupled NLS equation hierarchy, as a nonlinear Hermitian system, displays far richer nonlinear dynamics than the single-component case, and admits a more diverse set of vector localized wave solutions \cite{zhaolax,experiment4,db2,db8,dd1,dbvs}. The established quantitative correspondence between nonlinear Hermitian systems and linear non-Hermitian models in the scalar case motivates us to investigate potentially more intricate relationships in the vector case. Without loss of generality for nonlinear coupled systems, we choose a two-component Manakov system to demonstrate their intrinsic relationships,  which can be written as follows \cite{cnse,zhaolax,manakov}:
\begin{subequations}\label{CNLS}
\begin{align}
i\frac{\partial\psi_1}{\partial \xi}+\frac{1}{2}\frac{\partial^2\psi_1}{\partial \tau^2}+\sigma(|\psi_1|^2+|\psi_2|^2)\psi_1=0,\\
i\frac{\partial\psi_2}{\partial \xi}+\frac{1}{2}\frac{\partial^2\psi_2}{\partial \tau^2}+\sigma(|\psi_1|^2+|\psi_2|^2)\psi_2=0.
\end{align}
\end{subequations}
The nonlinear coupled model admits the Lax pair given by Eq.\eqref{spectral problems} with \cite{zhaolax}:
\begin{subequations}\label{CNLS UV}
\begin{align}
&U=i(\frac{\lambda}{2}(\sigma_3+\textbf{I})+Q),\\
&V=i(\frac{\lambda^2}{4}(\sigma_3+\textbf{I})+\frac{\lambda}{2}Q-\frac{1}{2}\sigma_3(Q^2+iQ_x)+s\textbf{I}),
\end{align}
\end{subequations}
and
\begin{eqnarray*}
Q&=&\begin{pmatrix}
    0 & \sigma\psi_1^* & \sigma\psi_2^*\\
\psi_1 & 0 & 0\\
\psi_2 & 0 & 0\\
\end{pmatrix},
\sigma_3=\begin{pmatrix}
    1 & 0 & 0\\
0 & -1 & 0\\
0 & 0 & -1\\
\end{pmatrix},
\end{eqnarray*}
where $\textbf{I}$ is the $3 \times 3$ identity matrix.  $s = s_1^2 + s_2^2$ is a real constant, and  $\lambda $ is complex spectral parameter. It is evident that $U$ and $V$ are non-Hermitian matrices, since  $U \neq U^\dagger$ and $V \neq V^\dagger$. Extending the analytical framework established for the scalar NLS system, we first examine the eigenvalue degeneracy of the Lax pair under a vector plane wave background with $\psi_{1}=s_1 e^{i(s\sigma - k^2/2)\xi + \textrm{i}k\tau}$ and $\psi_{2} = s_2 e^{i(s\sigma - k^2/2)\xi - \textrm{i}k\tau}$. The parameter $k$ denotes the frequency of the background. By introducing a transformation matrix $P = \mathrm{diag}(1,e^{-i(s^2 \sigma\, \xi + \frac{k^2}{2}\xi - k\tau)},e^{i(s^2 \sigma\, \xi + \frac{k^2}{2}\xi - k\tau)})$,  the variable-coefficient differential equation Eq. \eqref{spectral problems}  with  Eq. \eqref{CNLS UV}  can be converted into a constant-coefficient equation:
\begin{subequations}\label{CNLS U_0V_0}
\begin{align}
 U_0=&\begin{pmatrix}\label{CNLS U_0V_0}
    i\lambda & is_1\sigma & is_2\sigma\\
is_1 & -ik & 0\\
is_2 & 0 & ik\\
\end{pmatrix},\\
V_0=&-i\frac{1}{2}U_0^2-is(\sigma-1)\textbf{I}.
\end{align}
\end{subequations}

\begin{figure}[t]
\begin{center}
\subfigure{\includegraphics[width=80mm]{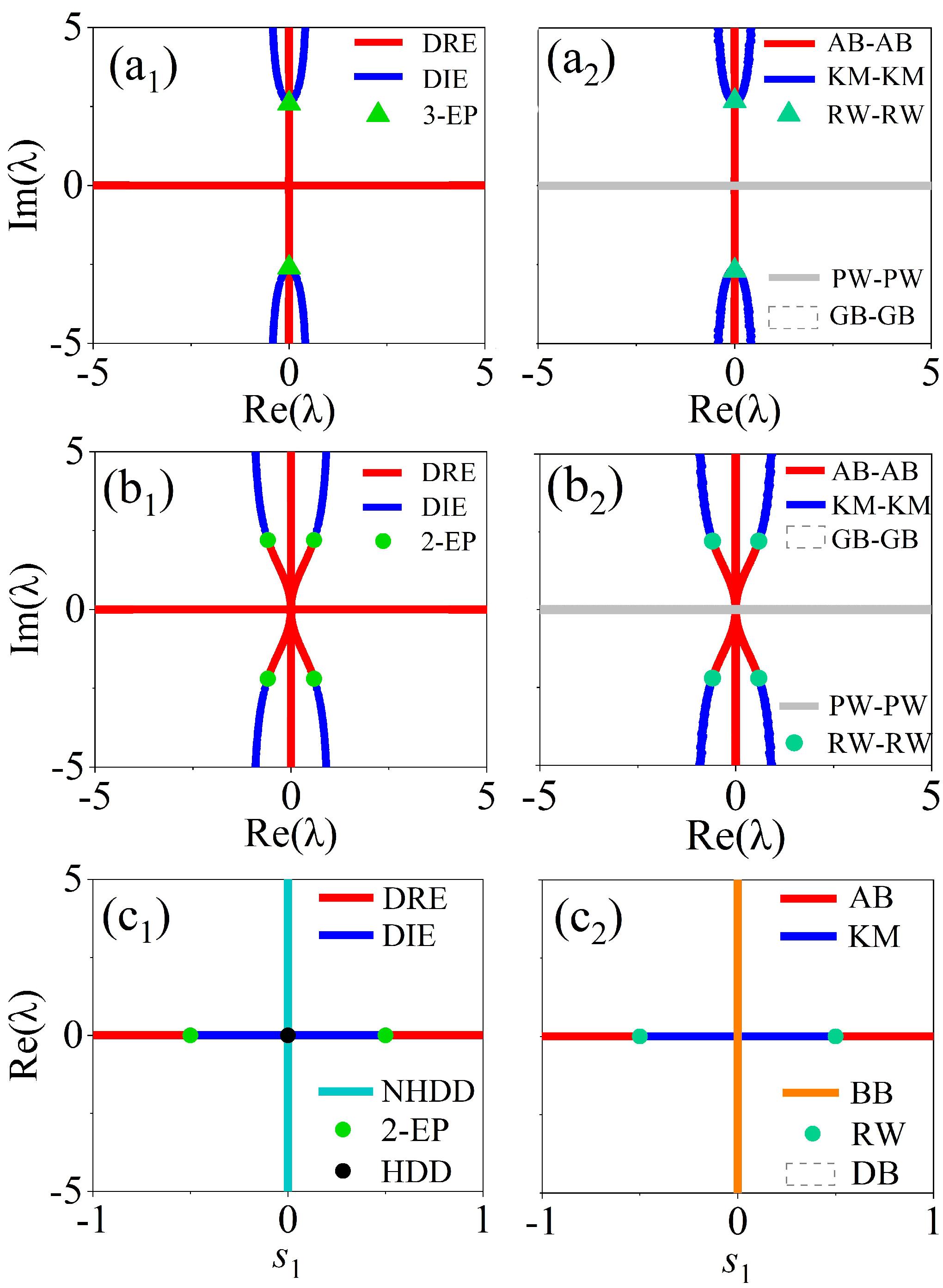}}
\end{center}
\caption{
The left column displays  eigenvalue degeneracies for $\sigma=1$. (a$_1$) and (b$_1$) show the degeneracy regions near the EPs in the Re$(\lambda)$-Im$(\lambda)$ plane for $k = 0.5$ and $k = 1$, respectively; (c$_1$) shows these regions in the $s_1$-Im$(\lambda)$ plane for $s_2 = 0$ and Im$(\lambda)=1$. The red curves, blue curves, green dots and green triangles represent DRE, DIE, 2-EPs and 3-EPs; the cyan line and black dot represent NHDD and HDD, respectively.
The right column presents the existence regions of localized wave solutions for $\sigma=1$: (a$_2$) $k = 0.5$ and  (b$_2$) $k = 1$. The red, blue, orange and gray curves represent AB-AB, KM-KM, BB and PW-PW solutions, respectively, with blank regions representing GB-GB.
Green triangles denote double RW-RW solutions, while green dots indicate single RW-RW solutions.
(c$_2$) The soliton existence regions in the $s_1$-Im$(\lambda)$ plane for $s_2 = 0$ and Im$(\lambda)=1$.
}\label{CNLSg1}
\end{figure}

\begin{figure}[t]
\begin{center}
\subfigure{\includegraphics[width=80mm]{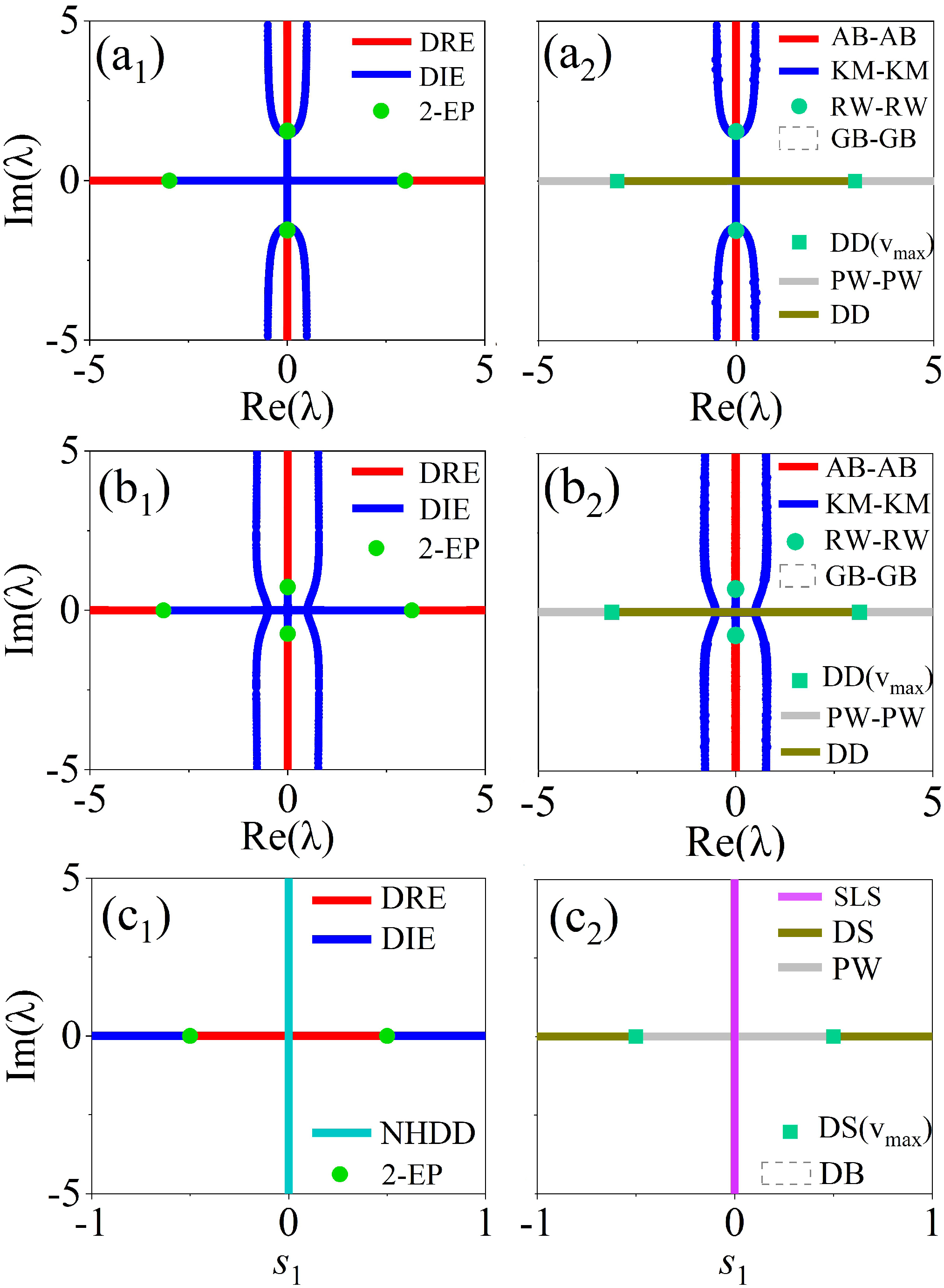}}
\end{center}
\caption{The left column shows eigenvalue degeneracies for $\sigma=-1$. (a$_1$) and (b$_1$) show the degeneracy regions near the EPs in the Re$(\lambda)$-Im$(\lambda)$ plane  for $k = 0.5$ and $k = 0.75$, respectively; (c$_1$) plots these regions in the $s_1$-Im$(\lambda)$ plane for $s_2 = 0$ and Re$(\lambda)=1$. The red curves, blue curves, green dots and cyan lines represent DRE, DIE, 2-EPs and NHDD, respectively. The right column illustrates  the existence regions of localized wave solutions for $\sigma=-1$:  (a$_2$) $k = 0.5$ and (b$_2$) $k = 0.75$. The red lines, blue curves,  gray circles  denote AB-AB, KM-KM and PW-PW solutions;  olive green lines, green squares and green circles represent  DD, DD solutions at maximum velocity and RW-RW solutions, respectively, with blank regions representing GB-GB.
(c$_2$) The soliton existence regions in the $s_1$-Im$(\lambda)$ plane for $s_2 = 0$ and Re$(\lambda)=1$.
}\label{CNLSg-1}
\end{figure}

As established in the preceding analysis, the matrix $U_0$ Eq.\eqref{CNLS U_0V_0} can be identified as the Hamiltonian of a linear non-Hermitian system, with eigenstates obeying the following eigenvalue equation:
\begin{eqnarray}\label{CNLS Ephi}
\begin{pmatrix}
    \lambda & is_1\sigma & is_2\sigma\\
is_1 & -ik & 0\\
is_2 & 0 & ik\\
\end{pmatrix}\varphi=E\varphi.
\end{eqnarray}
It is easy to get that there are three energy eigenvalues ($E_1$, $E_2$, and $E_3$).  We identify six distinct types of eigenvalue degeneracies of the real or imaginary spectrum of the linear non-Hermitian matrices, including: DRE, DIE, triple degeneracy in the real part, triple degeneracy in the imaginary part, double degeneracy, and triple degeneracy.

\begin{figure}[t]
\begin{center}
\subfigure{\includegraphics[width=87mm]{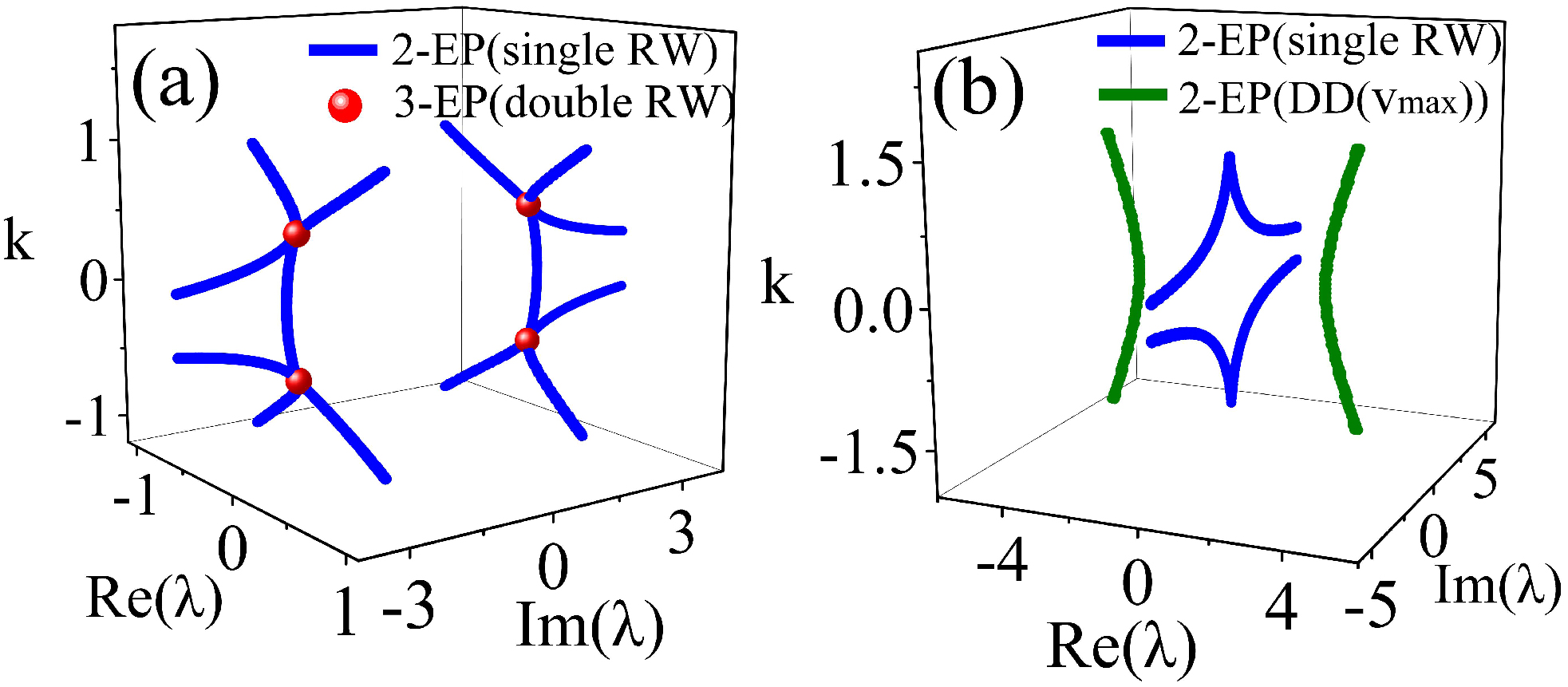}}
\end{center}
\caption{
Distribution of EPs in the three-dimensional parameter space of the coupled Manakov system: (a) $\sigma = 1$, (b) $\sigma = -1$.
The blue curves and red dots represent 2-EPs and 3-EPs, respectively, which correspond to the single and double RW-RW solutions.
3-EPs exist only in the case of $\sigma>0$.
The 2-EPs located on the imaginary axis (green curves) represent the DD solutions at maximum velocity.
Other parameters: $s_1 = s_2 = 1$.
}\label{CNLSep}
\end{figure}

We begin by investigating the eigenvalue degeneracies in the $\sigma=1$ case, as illustrated in left column of Fig.~\ref{CNLSg1}. The results reveal a richer degeneracy structure compared to the scalar case. On the plane wave background, in addition to DRE, DIE, and 2-EP, a third-order exceptional point (3-EP) degeneracy can also arise under certain conditions, where the eigenvectors of the three energy eigenvalue branches become identical and non-orthogonal, leading to three coalescing eigenstates. This case is demonstrated in  Fig.~\ref{CNLSg1}(a$_1$) with setting $s_1=s_2=1$ and $k=0.5$. The green triangles on the imaginary axis mark the 3-EPs, which exist exclusively when $|k| = 0.5$ (see the red points in Fig.~\ref{CNLSep}). In these points,  the non-Hermitian Hamiltonian $U_0$  is nondiagonalizable and admits only a Jordan canonical form.  The DRE regions correspond to points located on both the real and imaginary axes (excluding the 3-EPs), whereas the DIE regions are situated off the axes, emerging as two branches symmetrically distributed about the real axis. In the general case, the system exhibits DRE, DIE, and 2-EP, as shown in Fig.~\ref{CNLSg1}(b$_1$) with $k=1$.  The DRE regions are expanded, appearing not only along both axes but also extending beyond them. Notably, the four 2-EPs points no longer lie on the axes; instead, they partition the DRE and DIE regions in the Re($\lambda$)-Im($\lambda$) plane. More generally, we present the distribution regions of 2-EPs and 3-EPs in the three-dimensional case, using the (Re($\lambda$)-Im($\lambda$)-$k$) parameter space, as shown in Fig.~\ref{CNLSep}(a).

In contrast to the $\sigma=1$ case, the linear non-Hermitian system with $\sigma=-1$ (where $s_1=s_2=1$) exhibits DRE, DIE, and 2-EP characteristics but does not allow 3-EP.  Examples are illustrated in Figs.~\ref{CNLSg-1}(a$_1$) and (b$_1$) for $k=0.5$ and $k=0.75$, respectively. DRE regions are confined to the real and imaginary axes, in contrast to DIE regions which occur both on-axis and as symmetric off-axis distributions. In this regime, the system exhibits four 2-EPs: two located on the real axis and two on the imaginary axis. Particularly,  two 2-EPs on the imaginary axis approach each other as the frequency $k$ increases. In the limit $k \rightarrow \infty$, these two 2-EPs coalesce into a single 3-EP at $\lambda = 0$, as shown in Fig.~\ref{CNLSep}(b).

In the coupled system, when the amplitude of one of the plane wave backgrounds is set to zero (here we choose $s_2=0$), the non-Hermitian Hamiltonian Eq.~\eqref{CNLS Ephi} reduces to that of a scalar system, as discussed in the previous section. Consequently, the degeneracy regions coincide with those shown in Figs.~\ref{NLSg1} and \ref{NLSg-1}. Here, we further analyze the eigenvalue degeneracies as the amplitude $s_1$, as shown in Fig.~\ref{CNLSg1}(c$_1$) and Fig.~\ref{CNLSg-1}(c$_1$). We observe that when $s_1=1/2, s_2=0$, two 2-EPs emerge at Re$(\lambda)=0$ ($\sigma=1$) or Im$(\lambda)=0$ ($\sigma=-1$). In addition, when $s_1 = s_2 = 0$, the non-Hermitian Hamiltonian is
$U_0=\begin{pmatrix}
    i\lambda & 0 & 0\\
0 & 0 & 0\\
0 & 0 & 0\\
\end{pmatrix}$, which gives rise to the eigenvalues $E_1=E_2=0,E_3=i\lambda$. Obviously, the eigenvalues $E_1$ and $E_2$ are identically zero, resulting in a double degeneracy state. When Re$(\lambda)=0$, we obtain $U_0 = U_0^\dagger$, indicating the system exhibits HDD, as shown in Fig.~\ref{CNLSg1}(c$_1$) with black circle dot. In contrast, when Re$(\lambda)\neq0$, $U_0$ remains non-Hermitian ($U_0 \neq U_0^\dagger$) and the system maintains non-Hermitian eigenvalue degeneracy, the eigenvectors retain linearly independent, allowing $U_0$ to be diagonalized. Therefore, this case corresponds to non-Hermitian double  degeneracy (NHDD) rather than EP, as depicted in Fig.~\ref{CNLSg1}(c$_1$) Fig.~\ref{CNLSg-1}(c$_1$) with cyan lines. At $\lambda = 0$, all three eigenvalues become degenerate, $U_0$ becomes a zero matrix, converting the originally non-Hermitian Hamiltonian into a Hermitian one. In this case, a Hermitian triple degeneracy emerges.

Using the framework developed for the scalar NLS system, we classify nonlinear localized waves in the coupled Hermitian system Eq.~\eqref{CNLS} across the parameter space of eigenvalue spectral degeneracies, thereby revealing a mapping between linear non-Hermitian matrix degeneracies and nonlinear Hermitian vector waves. By solving the Lax pair (Eq.\eqref{CNLS UV}) via the Darboux transformation, eight types of nonlinear localized wave solutions and vector plane wave solutions (PW-PW) can be constructed, including KM-KM \cite{zhaolax,kmkm1}, AB-AB \cite{zhaolax,abab}, GB-GB, RW-RW \cite{zhaolax,rwrw}, dark-bright solitons (DB) \cite{experiment4,db2,db8}, dark-dark solitons (DD) \cite{dd1,dd6}, bright-bright solitons (BB) \cite{bb3,bb6} and vector SLS.
For $s_1 = s_2 = 1$, the existence regions of these solutions under different relative frequency $k$ are illustrated in Fig.~\ref{CNLSg1}(a2)-(b2) (for $\sigma = 1$) and Fig.~\ref{CNLSg-1}(a$_2$)-(b$_2$) (for $\sigma = -1$).  Comparing these with Figs.~\ref{CNLSg1}(a$_1$)-(b$_1$) and~\ref{CNLSg-1}(a$_1$)-(b$_1$), we get that the KM-KM/DD and AB-AB/PW-PW solutions map to DIE and DRE, respectively, while the vector GB solutions still have no degeneracies. It should be emphasized that the RW-RW solution in Fig.~\ref{CNLSg1}(a$_2$) represents double RWs, corresponding to 3-EPs in the non-Hermitian system, strikingly different from other RW-RW solutions that correspond to 2-EPs, analogous to the scalar case. In the MS regime, the DD solutions with maximum velocity are mapped to 2-EPs,  but their velocities are generally below the system sound speed ($c_s =\sqrt{2}$) in Bose-Einstein condensate systems, strikingly distinct from the scalar system. Only when the frequency vanishes does the maximum velocity attain $c_s$ \cite{dd6}. For the case $s_2 = 0$, the existence regions of various solutions are shown in Fig.~\ref{CNLSg1}(c$_2$) and Fig.~\ref{CNLSg-1}(c$_2$) for $\sigma = 1$ and $\sigma = -1$, respectively. Here, DB solutions occupy the blank regions, indicating they do not belong to any degeneracy spectrum.  When $s_1 = s_2 = 0$, BB solutions (for $g > 0$) and vector SLS solutions (for $g < 0$) are NHDD.

The above analysis demonstrates that coupled nonlinear systems admit significantly more abundant degeneracies in the real or imaginary spectra of linear non-Hermitian matrices. In particular, the coupled system's multi-band Hamiltonian enables richer EP, as exemplified by the 3-EPs case shown in Fig.~\ref{CNLSg1}(a$_1$). Namely, it allows more than two bands to coalesce at a single point, forming higher-order EPs that have attracted substantial research interest in recent years \cite{hep2,hep3,hep5}. The $N$th-order EPs can only exist in an $M$-state Hamiltonian when $M \geq N$. As established, the scalar NLS system's $U_0$  admits only two eigenstates, restricting the system to 2-EPs formation. By contrast, the two-component  Manakov system's $U_0$ possesses three eigenstates, enabling the realization of 3-EPs.

To better visualize these EPs geometries, we treat the parameter $k$ in the non-Hermitian Hamiltonian as an additional dimension. This approach enables construction of three-dimensional parameter-space slices for $\sigma= \pm 1$, as shown in Fig.~\ref{CNLSep}. It reveals that 2-EPs distributions constitute continuous exceptional arcs. For $\sigma = 1$ case, the exceptional arcs  (blue curves in Fig.~\ref{CNLSep}(a)) correspond to single RW-RW solutions in the nonlinear Hermitian system. For $\sigma = -1$, the 2-EPs located on the imaginary axis (green curves in Fig.~\ref{CNLSep}(b)) correspond to the DD solutions at maximum velocity, while other 2-EPs (blue curves in Fig.~\ref{CNLSep}(b)) represent single RW-RW solutions. The merging of two 2-EPs branches from different eigenvalues causes three eigenvalues to coincide at their intersection point, creating a 3-EP  (red circles in Fig.~\ref{CNLSep}(a)). It is worth noting that 3-EPs exist only for $\sigma > 0$, while for $\sigma < 0$ they can arise only at $k \rightarrow \infty$. At a 3-EP, all three eigenvalues and their corresponding eigenvectors coalesce, reducing the rank of $U_0$ to one and rendering it non-diagonalizable. Then, the parameters corresponding to 3-EPs in the non-Hermitian model generate two RWs in the nonlinear Hermitian system.

\section{conclusion}\label{conclusion}

We establish a correspondence between nonlinear Hermitian systems and linear non-Hermitian models through analysis of nonlinear localized wave solutions and spectral degeneracies of linear non-Hermitian matrices, with choosing the NLS equation integrability hierarchy.  Our results demonstrate that,  in both scalar and vector cases, KM and DD solutions correspond to DIE, while AB and PW solutions correspond to DRE; BS and SLS solutions relate to NHDD;  2-EPs map to single RW solutions in the modulational instability regime and to DS with maximum velocity in the stability regime. Particularly, 3-EPs emerge in two-component coupled systems, corresponding to vector solutions with two RWs. These findings provide a novel perspective for understanding nonlinear localized waves and may be extended to other nonlinear systems with higher-order effects.

\section*{Acknowledgments}
L.-C. Zhao was supported by the National Natural Science Foundation of China (Contract No. 12375005, 12235007, 12247103). Y.-H. Qin was supported by the National Natural Science Foundation of China (Grant No. 12405004), the Natural Science Foun dation of Xinjiang Uygur Autonomous Region Project (Grant No. 2024D01C232), the Scientific Research Projects Funded by the Basic Research Business Expenses of Autonomous Region Universities (Grant No. XJEDU2024P011), and the Program of ``Tianchi Talent" Introduction Plan in Xinjiang Uygur Autonomous Region.


\begin{thebibliography}{99}
\bibitem{book1} N. Moiseyev, \textit{Non-Hermitian Quantum Mechanics},
\newblock{Cambridge University Press (2011)}.

\bibitem{book2} C. M. Bender, P. E. Dorey, C. Dunning, A. Fring, D. W. Hook, H. F. Jones, S. Kuzhel, G. L\'evai, and R. Tateo, \textit{PT Symmetry: In Quantum and Classical Physics},
\newblock{World Scientific Publishing (2019)}.

\bibitem{book3} F. Bagarello and C. Trapani, \textit{Non-Hermitian Hamiltonians in Quantum Physics},
\newblock{Springer, New York (2016)}.

\bibitem{rmp1} E. J. Bergholtz, J. C. Budich, and F. K. Kunst, Exceptional topology of non-Hermitian systems,
\newblock\href{https://doi.org/10.1103/RevModPhys.93.015005}{Rev. Mod. Phys. {\bf93}, 015005 (2021)}.

\bibitem{ap1} Y. Ashida, Z. Gong, and M. Ueda, Non-Hermitian physics,
\newblock\href{https://doi.org/10.1080/00018732.2021.1876991}{Adv. Phys. {\bf69}, 249 (2021)}.

\bibitem{prl2} N. Okuma, K. Kawabata, K. Shiozaki, S. Ryu, and M. Sato, Topological origin of non-Hermitian skin effects,
\newblock\href{https://doi.org/10.1103/PhysRevLett.124.086801}{Phys. Rev. Lett. {\bf124}, 086801 (2020)}.

\bibitem{np1} T. Helbig, T. Hofmann, S. Imhof, C. Nolte, M. Br\"uning, T.-E. Lee, A. Szameit, L. W. Molenkamp, and F. Schindler, Generalized bulk-boundary correspondence in non-Hermitian topolectrical circuits,
\newblock\href{https://doi.org/10.1038/s41567-020-0922-9}{Nat. Phys. {\bf16}, 747 (2020)}.

\bibitem{np2} H. Zhou, C. Peng, Y. Yoon, C. W. Hsu, K. A. Nelson, L. Fu, J. D. Joannopoulos, M. Solja\v{c}i\'{c}, and B. Zhen, Observation of bulk Fermi arc and polarization half charge from paired exceptional points,
\newblock\href{https://doi.org/10.1126/science.aap9859}{Science {\bf359}, 1009 (2018)}.

\bibitem{sc1} W. Chen, \c{S}. Kaya\"{O}zdemir, G. Zhao, J. Wiersig, and L. Yang, Exceptional points enhance sensing in an optical microcavity,
\newblock\href{https://doi.org/10.1038/nature23281}{Nature {\bf548}, 192-196 (2017)}.

\bibitem{nature1} S. Yao and Z. Wang, Edge states and topological invariants of non-Hermitian systems,
\newblock\href{https://doi.org/10.1103/PhysRevLett.121.086803}{Phys. Rev. Lett. {\bf121}, 086803 (2018)}.

\bibitem{prl1} R. El-Ganainy, K. G. Makris, M. Khajavikhan, Z. H. Musslimani, S. Rotter, and D. N. Christodoulides, Non-Hermitian physics and PT symmetry,
\newblock\href{https://doi.org/10.1038/nphys4323}{Nat. Phys. {\bf14}, 11 (2018)}.

\bibitem{zhao20} L.-C. Zhao, W. Wang, Q. Tang, Z.-Y. Yang, W.-L. Yang, and J. Liu, Spin soliton with a negative-positive mass transition,
\newblock\href{https://doi.org/10.1103/PhysRevA.101.043621}{Phys. Rev. A {\bf101}, 043621 (2020)}.

\bibitem{meng22} L. Z. Meng, S. W. Guan, and L.-C. Zhao, Negative mass effects of a spin soliton in Bose-Einstein condensates,
\newblock\href{https://doi.org/10.1103/PhysRevA.105.013303}{Phys. Rev. A {\bf105}, 013303 (2022)}.

\bibitem{gaox25} X. Gao, L. Z. Meng, and L.-C. Zhao, Dark-bright solitons with positive mass in Manakov cases with repulsive interactions,
\newblock\href{https://doi.org/10.1103/PhysRevE.111.054209}{Phys. Rev. E {\bf111}, 054209 (2025)}.

\bibitem{darboux1} V. B. Matveev and M. A. Salle, \textit{Darboux Transformations and Solitons},
\newblock{Springer, Berlin (1991)}.

\bibitem{darboux2} C. Gu, H. Hu, and Z. Zhou, \textit{Darboux Transformations in Integrable Systems: Theory and Their Applications to Geometry},
\newblock{Springer, Dordrecht (2004)}.

\bibitem{JYang} J. K. Yang, \textit{Nonlinear Waves in Integrable and Nonintegrable Systems},
\newblock{Society for Industrial and Applied Mathematics, Philadelphia (2010)}.

\bibitem{ds1} A. Shabat and V. Zakharov, Exact theory of two-dimensional self-focusing and one-dimensional self-modulation of waves in nonlinear media,
\newblock{Sov. Phys. JETP {\bf34}, 62 (1972)}.

\bibitem{ds2} V. E. Zakharov and A. B. Shabat, Interaction between solitons in a stable medium,
\newblock{Sov. Phys. JETP {\bf37}, 823 (1973)}.

\bibitem{zhaolax} L. Ling and L.-C. Zhao, Modulational instability and homoclinic orbit solutions in vector nonlinear Schr\"odinger equation,
\newblock\href{https://doi.org/10.1016/j.cnsns.2019.01.008}{Commun. Nonlinear Sci. Numer. Simul. {\bf72}, 449-471 (2019)}.

\bibitem{dd6} L. Ling, L.-C. Zhao, and B. Guo, Darboux transformation and multi-dark soliton for N-component nonlinear Schr\"odinger equations,
\newblock\href{https://doi.org/10.1088/0951-7715/28/9/3243}{Nonlinearity {\bf28}, 3243 (2015)}.

\bibitem{cp1} J.-W. Ryu, J.-H. Han, C.-H. Yi, M. J. Park, and H. C. Park, Exceptional classifications of non-Hermitian systems,
\newblock\href{https://doi.org/10.1038/s42005-024-01595-9}{Commun. Phys. {\bf7}, 109 (2024)}.

\bibitem{np4} K. Ochkan, R. Chaturvedi, V. K\"onye, L. Veyrat, R. Giraud, D. Mailly, A. Cavanna, U. Gennser, E. M. Hankiewicz, B. B\"uchner, J. van den Brink, and I. C. Fulga, Observation of non-Hermitian topology in a multi-terminal quantum Hall device,
\newblock\href{https://doi.org/10.1038/s41567-023-02337-4}{Nat. Phys. {\bf20}, 395-401 (2024)}.

\bibitem{nrp1} K. Ding, C. Fang, and G. Ma, Non-Hermitian topology and exceptional-point geometries,
\newblock\href{https://doi.org/10.1038/s42254-022-00516-5}{Nat. Rev. Phys. {\bf4}, 745-760 (2022)}.

\bibitem{BECbook} P. G. Kevrekidis, D. J. Frantzeskakis, and R. Carretero-Gonz\'alez, \textit{Emergent Nonlinear Phenomena in Bose-Einstein Condensates: Theory and Experiment},
\newblock{Springer, London (2008)}.

\bibitem{nonliearopbook} G. P. Agrawal, \textit{Nonlinear Fiber Optics}, 5th ed.,
\newblock{Academic Press, Boston (2012)}.

\bibitem{financialbook} W. A. Brock, D. A. Hsieh, and B. D. LeBaron, \textit{Nonlinear Dynamics, Chaos, and Instability: Statistical Theory and Economic Evidence},
\newblock{MIT Press, Cambridge, MA (1991)}.

\bibitem{experiment1} B. Kibler, J. Fatome, C. Finot, G. Millot, F. Dias, G. Genty, N. Akhmediev, and J. M. Dudley, The Peregrine soliton in nonlinear fibre optics,
\newblock\href{https://doi.org/10.1038/nphys1740}{Nat. Phys. {\bf6}, 790-795 (2010)}.

\bibitem{experiment2} A. Chabchoub, N. P. Hoffmann, and N. Akhmediev, Rogue wave observation in a water wave tank,
\newblock\href{https://doi.org/10.1103/PhysRevLett.106.204502}{Phys. Rev. Lett. {\bf106}, 204502 (2011)}.

\bibitem{experiment3} H. Bailung, S. K. Sharma, and Y. Nakamura, Observation of Peregrine solitons in a multicomponent plasma with negative ions,
\newblock\href{https://doi.org/10.1103/PhysRevLett.107.255005}{Phys. Rev. Lett. {\bf107}, 255005 (2011)}.

\bibitem{experiment4} C. Becker, S. Stellmer, P. Soltan-Panahi, S. D\"{o}rscher, M. Baumert, E.-M. Richter, J. Kronj\"{a}ger, K. Bongs, and K. Sengstock, Oscillations and interactions of dark and dar-bright solitons in Bose-Einstein condensates,
\newblock\href{https://doi.org/10.1038/nphys962}{Nat. Phys. {\bf4}, 496-501 (2008)}.

\bibitem{kmexp} B. Kibler, J. Fatome, C. Finot, G. Le Bihan, P. Morin, L. Provino, N. Akhmediev, Iliya Kibler (dup), D. Huard, and G. Millot, Observation of Kuznetsov-Ma soliton dynamics in optical fibre,
\newblock\href{https://doi.org/10.1038/srep00463}{Sci. Rep. {\bf2}, 463 (2012)}.

\bibitem{abexp1} J. M. Dudley, G. Genty, F. Dias, B. Kibler, and N. Akhmediev, Modulation instability, Akhmediev Breathers and continuous wave supercontinuum generation,
\newblock\href{https://doi.org/10.1364/OE.17.021497}{Opt. Express {\bf17}, 21497 (2009)}.

\bibitem{abexp2} K. Hammani, B. Wetzel, B. Kibler, N. Akhmediev, and Y. Koonath, Spectral dynamics of modulation instability described using Akhmediev breather theory,
\newblock\href{https://doi.org/10.1364/OL.36.002140}{Opt. Lett. {\bf36}, 2140 (2011)}.

\bibitem{rwexp} D. R. Solli, C. Ropers, P. Koonath, and B. Jalali, Optical rogue waves,
\newblock\href{https://doi.org/10.1038/nature06402}{Nature {\bf450}, 1054 (2007)}.

\bibitem{dsexp} A. M. Weiner, J. P. Heritage, R. J. Hawkins, J. Pietras, and J. P. Gordon, Experimental observation of the fundamental dark soliton in optical fibers,
\newblock\href{https://doi.org/10.1103/PhysRevLett.61.2445}{Phys. Rev. Lett. {\bf61}, 2445 (1988)}.

\bibitem{bsexp} L. F. Mollenauer, R. H. Stolen, and J. P. Gordon, Experimental observation of picosecond pulse narrowing and solitons in optical fibers,
\newblock\href{https://doi.org/10.1103/PhysRevLett.45.1095}{Phys. Rev. Lett. {\bf45}, 1095 (1980)}.

\bibitem{nos1} G. P. Agrawal, \textit{Applications of Nonlinear Fiber Optics}, 3rd ed.,
\newblock{Academic Press, San Diego (2020)}.

\bibitem{nos2} Y. V. Kartashov, B. A. Malomed, and L. Torner, Solitons in nonlinear lattices,
\newblock\href{https://doi.org/10.1103/RevModPhys.83.247}{Rev. Mod. Phys. {\bf83}, 247 (2011)}.

\bibitem{nos3} J. M. Dudley, G. Genty, and S. Coen, Supercontinuum generation in photonic crystal fiber,
\newblock\href{https://doi.org/10.1103/RevModPhys.78.1135}{Rev. Mod. Phys. {\bf78}, 1135 (2006)}.

\bibitem{bec} Y. V. Bludov, V. V. Konotop, and N. Akhmediev, Matter rogue waves,
\newblock\href{https://doi.org/10.1103/PhysRevA.80.033610}{Phys. Rev. A {\bf80}, 033610 (2009)}.

\bibitem{darboux5} L.-C. Zhao and J. Liu, Localized nonlinear waves in a two-mode nonlinear fiber,
\newblock\href{https://doi.org/10.1364/JOSAB.29.003119}{J. Opt. Soc. Am. B {\bf29}, 3119 (2012)}.

\bibitem{Hirota2} R. K. Bullough and P. J. Caudrey, \textit{Solitons},
\newblock{Springer, Berlin, Heidelberg (1980)}.

\bibitem{inverse} M. J. Ablowitz and H. Segur, \textit{Solitons and the Inverse Scattering Transform},
\newblock{SIAM, Philadelphia (1981)}.

\bibitem{NLSlax} B. Guo, L. Ling, and Q. P. Liu, Nonlinear Schr\"odinger equation: generalized Darboux transformation and rogue wave solutions,
\newblock\href{https://doi.org/10.1103/PhysRevE.85.026607}{Phys. Rev. E {\bf85}, 026607 (2012)}.

\bibitem{ep1} T. Kato, \textit{Perturbation Theory for Linear Operators},
\newblock{Springer, Berlin (2013)}.

\bibitem{ep2} W. D. Heiss and A. L. Sannino, Avoided level crossing and exceptional points,
\newblock\href{https://doi.org/10.1088/0305-4470/23/7/022}{J. Phys. A: Math. Gen. {\bf23}, 1167-1178 (1990)}.

\bibitem{km1} E. A. Kuznetsov, Solitons in a parametrically unstable plasma,
\newblock\href{https://doi.org/10.1070/PU1977v020n08ABEH004533}{Dokl. Akad. Nauk SSSR {\bf236}, 508-511 (1977)}.

\bibitem{km2} Y. C. Ma, The perturbed plane-wave solutions of the cubic Schr\"odinger equation,
\newblock\href{https://doi.org/10.1002/sapm197960143}{Stud. Appl. Math. {\bf60}, 43 (1979)}.

\bibitem{ab1} N. Akhmediev, V. M. Eleonskii, and N. E. Kulagin, Generation of periodic trains of picosecond pulses in an optical fiber: exact solutions,
\newblock{Sov. Phys. JETP {\bf62}, 894 (1985)}.

\bibitem{ab2} N. N. Akhmediev and V. I. Korneev, Modulation instability and periodic solutions of the nonlinear Schr\"odinger equation,
\newblock\href{https://doi.org/10.1007/BF01037866}{Theor. Math. Phys. {\bf69}, 1089 (1986)}.

\bibitem{ab3} N. N. Akhmediev, V. M. Eleonskii, and N. E. Kulagin, Exact first-order solutions of the nonlinear Schr\"odinger equation,
\newblock{Theor. Math. Phys. {\bf72}, 809 (1987)}.

\bibitem{gb} M. Tajiri and Y. Watanabe, Breather solutions to the focusing nonlinear Schr\"odinger equation,
\newblock\href{https://doi.org/10.1103/PhysRevE.57.3510}{Phys. Rev. E {\bf57}, 3510 (1998)}.

\bibitem{rw} D. H. Peregrine, Water waves, nonlinear Schr\"odinger equations and their solutions,
\newblock\href{https://doi.org/10.1017/S0334270000003891}{ANZIAM J. {\bf25}, 16-43 (1983)}.

\bibitem{dbkm} A. A. Gelash and V. E. Zakharov, Superregular solitonic solutions: a novel scenario for the nonlinear stage of modulation instability,
\newblock\href{https://doi.org/10.1088/0951-7715/27/4/R1}{Nonlinearity {\bf27}, R1 (2014)}.

\bibitem{dbab} N. N. Akhmediev, A. Ankiewicz, and M. Taki, Waves that appear from nowhere and disappear without a trace,
\newblock\href{https://doi.org/10.1016/j.physleta.2008.12.036}{Phys. Lett. A {\bf373}, 675-678 (2009)}.

\bibitem{dbrw1} N. Akhmediev, A. Ankiewicz, and J. M. Soto-Crespo, Rogue waves and rational solutions of the nonlinear Schr\"odinger equation,
\newblock\href{https://doi.org/10.1103/PhysRevE.80.026601}{Phys. Rev. E {\bf80}, 026601 (2009)}.

\bibitem{db2} T. Busch and J. R. Anglin, Dark-bright solitons in inhomogeneous Bose-Einstein condensates,
\newblock\href{https://doi.org/10.1103/PhysRevLett.87.010401}{Phys. Rev. Lett. {\bf87}, 010401 (2001)}.

\bibitem{db8} L. Ling and X. Sun, The stability of bright-dark solitons in defocusing coupled nonlinear Schr\"odinger equation,
\newblock\href{https://doi.org/10.1016/j.padiff.2022.100342}{Partial Differ. Equ. Appl. Math. {\bf5}, 100342 (2022)}.

\bibitem{dd1} P. \"Ohberg and L. Santos, Dark solitons in a two-component Bose-Einstein condensate,
\newblock\href{https://doi.org/10.1103/PhysRevLett.86.2918}{Phys. Rev. Lett. {\bf86}, 2918 (2001)}.

\bibitem{dbvs} L. Ling, L.-C. Zhao, and B. Guo, Darboux transformation and classification of solution for mixed coupled nonlinear Schr\"odinger equations,
\newblock\href{https://doi.org/10.1016/j.cnsns.2015.08.023}{Commun. Nonlinear Sci. Numer. Simul. {\bf32}, 285-304 (2016)}.

\bibitem{manakov} S. V. Manakov, On the theory of two-dimensional stationary self-focusing of electromagnetic waves,
\newblock{Sov. Phys. JETP {\bf38}, 248 (1974)}.

\bibitem{cnse} P. G. Kevrekidis and D. J. Frantzeskakis, Solitons in coupled nonlinear Schr\"odinger models: a survey of recent developments,
\newblock\href{https://doi.org/10.1016/j.revip.2016.07.002}{Rev. Phys. {\bf1}, 140-153 (2016)}.

\bibitem{kmkm1} W. J. Che and S. C. Chen and C. Liu and X. X. Liu and H. Pu, Nondegenerate Kuznetsov-Ma solitons of Manakov equations and their physical spectra,
\newblock\href{https://doi.org/10.1103/PhysRevA.105.043526}{Phys. Rev. A {\bf105}, 043526 (2022)}.

\bibitem{abab} S. C. Chen and C. Liu, Hidden Akhmediev breathers and vector modulation instability in the defocusing regime,
\newblock\href{https://doi.org/10.1016/j.physd.2022.133364}{Physica D {\bf438}, 133364 (2022)}.

\bibitem{rwrw} Y. H. Qin, L. Ling, and L.-C. Zhao, Optical rogue-wave patterns in coupled defocusing systems,
\newblock\href{https://doi.org/10.1103/PhysRevA.108.023519}{Phys. Rev. A {\bf108}, 023519 (2023)}.

\bibitem{bb3} X. X. Liu, H. Pu, B. Xiong, and J. H. Liu, Formation and transformation of vector solitons in two-species Bose-Einstein condensates with a tunable interaction,
\newblock\href{https://doi.org/10.1103/PhysRevA.79.013423}{Phys. Rev. A {\bf79}, 013423 (2009)}.

\bibitem{bb6} G. Csire, D. Schumayer, and B. Apagyi, Effect of scattering lengths on the dynamics of a two-component Bose-Einstein condensate,
\newblock\href{https://doi.org/10.1103/PhysRevA.82.063608}{Phys. Rev. A {\bf82}, 063608 (2010)}.

\bibitem{hep2} K. Ding, G. Ma, M. Xiao, J. Q. Zhang, and C. T. Chan, Emergence, coalescence, and topological properties of multiple exceptional points and their experimental realization,
\newblock\href{https://doi.org/10.1103/PhysRevX.6.021007}{Phys. Rev. X {\bf6}, 021007 (2016)}.

\bibitem{hep3} H. Hodaei, A. U. Hassan, S. Wittek, H. Garcia-Gracia, R. El-Ganainy, D. N. Christodoulides, and M. Khajavikhan, Enhanced sensitivity at higher-order exceptional points,
\newblock\href{https://doi.org/10.1038/nature23280}{Nature {\bf548}, 187 (2017)}.

\bibitem{hep5} S. Wang, B. Hou, W. Lu, Y. Chen, Z. Q. Zhang, and C. T. Chan, Arbitrary order exceptional point induced by photonic spin-orbit interaction in coupled resonators,
\newblock\href{https://doi.org/10.1038/s41467-019-08826-6}{Nat. Commun. {\bf10}, 832 (2019)}.

\end{thebibliography}
\end{document}